\documentstyle[12pt]{article}
\begin{document} 
\global\parskip 6pt
\newcommand{\be}{\begin{equation}}
\newcommand{\ee}{\end{equation}}
\newcommand{\bea}{\begin{eqnarray}}
\newcommand{\eea}{\end{eqnarray}}
\newcommand{\non}{\nonumber}

\begin{titlepage}
\hfill{hep-th/9801145}
\vspace*{1cm}
\begin{center}
{\Large\bf String Theory Formulation of\\
anti-de Sitter Black Holes}\\
\vspace*{2cm}
Danny Birmingham\footnote{E-mail: dannyb@ollamh.ucd.ie}\\
\vspace*{.5cm}
{\em Department of Mathematical Physics\\
University College Dublin\\
Belfield, Dublin 4, Ireland}\\
\vspace{2cm}
\begin{abstract}
It is shown that the five-dimensional anti-de Sitter black
hole is a supersymmetric solution of the low-energy field
equations of type IIB string theory compactified on an
Einstein space. A statistical interpretation of the mass
dependence of the entropy can be obtained from
considerations of the three-dimensional BTZ black hole.
\end{abstract}
\vspace{1cm}
January 1998
\end{center}
\end{titlepage}

\section{Introduction}
The study of black holes in string theory has been the source
of recent developments in our understanding of the origin
of the Bekenstein-Hawking entropy formula \cite{SV}-\cite{Horowitz}.
In many cases, an important role is played by supersymmetry,
and the observation that D-branes carry charge for
the RR fields of string theory \cite{Polchinski}.
In three dimensions, a particularly interesting
example of a black hole, known as the BTZ black hole,
has been constructed  \cite{BTZ,BHTZ}, see \cite{Carlip} for a review.
The construction is based upon
the observation that by performing a quotient of three-dimensional
anti-de Sitter space ($\mathrm{adS}_{3}$),
one obtains a spacetime with the properties
of a black hole.
It was observed recently \cite{Strominger2}
that the microscopic entropy can
be understood for certain black holes whose near-horizon geometry
is locally equivalent to $\mathrm{adS}_{3}$.
This observation is based on the fact that $\mathrm{adS}_{3}$
has an asymptotic symmetry algebra consisting of left
and right Virasoro algebras \cite{Brown,CH}. In particular,
the entropy of the extreme BTZ black hole, viewed  as a
supersymmetric solution of heterotic string theory
without the use of RR  fields, was computed in \cite{BSS}.

In \cite{Banados}, a higher-dimensional generalization
of the BTZ construction was provided. The essential idea is
to take a quotient of anti-de Sitter space, yielding
a black hole with topology ${\mathbf{R}}^{d-1}\times S^{1}$, and
an explicit construction in five dimensions was presented.
An interesting aspect of these higher-dimensional black holes
is that the horizon is a circle, and thus one might hope
to be able to understand their entropy from a lower-dimensional point
of view. We show that the five-dimensional black hole
is a supersymmetric solution of type IIB string theory
compactified to $\mathrm{adS}_{5}\times K_{5}$, where $K_{5}$
is a compact internal Einstein space,
with only the field strength for the
RR $4$-form excited. At extremality, the black hole
has a non-zero horizon length, proportional to the square root
of the mass. A statistical interpretation of this mass
dependence can be achieved by relating it to the entropy
of the three-dimensional BTZ black hole.

\section{Compactification of Type IIB String Theory
on $\mathbf{adS_{5}}$}
The five-dimensional anti-de Sitter black hole is obtained
as a quotient of $\mathrm{adS}_{5}$, to which it is
locally equivalent \cite{Banados}.
To obtain this five-dimensional black
hole as a supersymmetric solution of type IIB string
theory, we seek a compactification to $\mathrm{adS}_{5}\times K_{5}$,
where $K_{5}$ is a compact internal space. The black hole
is then obtained by performing the necessary identifications.
We write the
$10$-dimensional coordinates as $x^{M} = (x^{\mu},y^{m})$,
with $\mu=0,1,2,3,4$, and $m=5,6,7,8,9$, and we
follow the conventions of \cite{Duff2}.
The covariant field equations for type IIB string theory
have been determined in \cite{Schwarz}. As observed in \cite{Schwarz},
the desired compactification can be obtained by setting
all the bosonic fields to zero, except for the metric
and anti-self-dual $5$-form field strength.
In this case, the only non-trivial equation of
motion is given by
\bea
R_{MN} = \frac{1}{4^{2}\times 6}
F_{MN_{1}N_{2}N_{3}N_{4}}F_{N}^{\phantom{N}
N_{1}N_{2}N_{3}N_{4}},
\label{eq1}
\eea
where the $5$-form is taken to be anti-self-dual. Thus, we have
\bea
F_{M_{1}M_{2}M_{3}M_{4}M_{5}} = - \frac{1}{5!}
\epsilon_{M_{1}M_{2}M_{3}M_{4}
M_{5}}^{\phantom{M_{1}M_{2}M_{3}M_{4}M_{5}}
N_{1}N_{2}N_{3}N_{4}N_{5}}F_{N_{1}N_{2}N_{3}N_{4}N_{5}},
\eea
with $\epsilon^{0123456789} = 1/\sqrt{-g}$.
We wish to obtain a product metric of the form
\bea
g_{\mu\nu} = g_{\mu\nu}(x),\;\;g_{mn}=g_{mn}(y),\;\;g_{\mu m}=0.
\eea
The required solution follows by choosing the ansatz \cite{Schwarz}
\bea
F_{\mu_{1}\mu_{2}\mu_{3}\mu_{4}\mu_{5}} &=&
Q \epsilon_{\mu_{1}\mu_{2}
\mu_{3}\mu_{4}\mu_{5}}, \non\\
F_{m_{1}m_{2}m_{3}m_{4}m_{5}} &=& -Q \epsilon_{m_{1}m_{2}
m_{3}m_{4}m_{5}}.
\eea
It is then straightforward to see that
\bea
R_{\mu\nu} &=& -\frac{Q^{2}}{4}g_{\mu\nu},\non\\
R_{mn} &=& \frac{Q^{2}}{4}g_{mn},
\label{curv}
\eea
with $R_{\mu m}=0$, and we note that the $10$-dimensional Ricci scalar
vanishes.

In order to establish the supersymmetry of the solution,
we must show that the Killing spinor equations
are satisfied.
Namely, we must show that
the supersymmetry variations of the fermionic fields
vanish in the compactified background.
The relevant terms in the supersymmetry transformations take
the form \cite{Duff2,Schwarz}
\bea
\delta{\psi}_{M} &=& \nabla_{M}\epsilon
+\frac{i}{4\times 480}\Gamma^{M_{1}M_{2}M_{3}M_{4}M_{5}}
\Gamma_{M}\epsilon
F_{M_{1}M_{2}M_{3}M_{4}M_{5}}\non\\
&+& \frac{1}{96}\left(
\Gamma_{M}^{{\phantom M}M_{1}M_{2}M_{3}} -
9\delta_{M}^{\phantom{M}M_{1}}\Gamma^{M_{2}M_{3}}\right)
\epsilon^{*}F_{M_{1}M_{2}M_{3}},
\eea
\bea
\delta \lambda = i\Gamma^{M}\epsilon^{*} \left(\frac{\nabla_{M}\phi}
{1-\phi^{*}\phi}\right) -\frac{i}{24}
\Gamma^{M_{1}M_{2}M_{3}}\epsilon F_{M_{1}M_{2}M_{3}}.
\eea
A representation of the Dirac
matrices which is relevant to the
$5+5$ split is given in \cite{Wetterich,Schwarz2}.
The $10$-dimensional Dirac matrices are denoted by $\Gamma^{A}$
and satisfy
\bea
\{\Gamma^{A},\Gamma^{B}\} = 2\eta^{AB},
\eea
with signature $\eta^{AB} = (-+\cdots+)$. The representation
is given by
\bea
\Gamma^{A} = (\Gamma^{\alpha},\Gamma^{a}) = (\gamma^{\alpha}
\otimes {\mathbf{1}}_{4}, \gamma^{6}\otimes \Sigma^{a}),
\eea
where the spacetime matrices are $8$-dimensional, and the
internal matrices are $4$-dimensional, satisfying
\bea
\{\gamma^{\alpha},\gamma^{\beta}\} = 2\eta^{\alpha\beta},\;\;
\{\Sigma^{a},\Sigma^{b}\} = 2\delta^{ab},
\eea
with signature $\eta^{\alpha\beta} = (-++++)$.
The $\gamma^{6}$ matrix is given explicitly by
\bea
\gamma^{6} = \left( \begin{array}{cc}
{\mathbf{1}}_{4}&0\\
0&-{\mathbf{1}}_{4} \end{array}
\right ),
\eea
and satisfies
\bea
\{\gamma^{6},\gamma^{\alpha}\} = 0,
\;\; (\gamma^{6})^{2} = {\mathbf{1}}_{8}.
\eea
The spacetime Dirac matrices can be written as \cite{Wetterich,Schwarz2}
\bea
\gamma^{0} &=& \left( \begin{array}{cc}
0&i\sigma^{1}\otimes{\mathbf{1}}_{2}\\
i\sigma^{1}\otimes{\mathbf{1}}_{2}&0 \end{array}
\right ),\;\;
\gamma^{1} = \left( \begin{array}{cc}
0&\sigma^{3}\otimes {\mathbf{1}}_{2}\\
\sigma^{3}\otimes{\mathbf{1}}_{2}&0 \end{array}
\right ),\;\;\non\\
\gamma^{2} &=& \left( \begin{array}{cc}
0&\sigma^{2}\otimes\tau^{2}\\
\sigma^{2}\otimes \tau^{2}&0 \end{array}
\right ),\;\;
\gamma^{3} = \left( \begin{array}{cc}
0&\sigma^{2}\otimes\tau^{1}\\
\sigma^{2}\otimes \tau^{1}&0 \end{array}
\right ),\;\;\non\\
\gamma^{4} &=& \left( \begin{array}{cc}
0&\sigma^{2}\otimes\tau^{3}\\
\sigma^{2}\otimes \tau^{3}&0 \end{array}
\right ),
\eea
where $\sigma^{i}$ and $\tau^{i}$ are the Pauli matrices.
The internal matrices are
\bea
\Sigma^{5} &=& \sigma^{1}\otimes{\mathbf{1}}_{2},\;\; \Sigma^{6} =
\sigma^{3}\otimes{\mathbf{1}}_{2},\;\; \Sigma^{7} = \sigma^{2}\otimes
\tau^{2},\non\\
\Sigma^{8} &=& \sigma^{2}\otimes \tau^{1},\;\; \Sigma^{9} =
\sigma^{2}\otimes \tau^{3},
\eea
with
\bea
\Gamma^{\alpha\beta} = \gamma^{\alpha\beta}\otimes
{\mathbf{1}}_{4},\;\;
\Gamma^{ab} = {\mathbf{1}}_{8}\otimes \Sigma^{ab}.
\eea
A matrix which also enters the analysis is defined by
\bea
J = \left( \begin{array}{cc}
0&{\mathbf{1}}_{4}\\
{\mathbf{1}}_{4}&0 \end{array}
\right ),
\eea
and satisfies $[J,\gamma^{\alpha}] = 0$.

We write the $10$-dimensional spinor
as $\epsilon(x,y) = \eta(x)\otimes\chi(y)$.
The Killing spinor equations then become
\bea
\delta \psi_{\mu} &=& 0 \Rightarrow\;\; \nabla_{\mu}\eta= \frac{Q}{8}
J\gamma_{\mu}\eta,   \non\\
\delta \psi_{m} &=& 0 \Rightarrow\;\;
\nabla_{m}\chi = - i\frac{Q}{8}\Sigma_{m}\chi,
\label{susy}
\eea
with the constraint $iJ\gamma^{6}\eta = \eta$.
The integrability conditions implied by
(\ref{susy}) are then precisely the conditions
(\ref{curv}). Thus, we have obtained $\mathrm{adS}_{5} \times K_{5}$
as a supersymmetric solution of the low-energy equations of
motion of type IIB string theory.
The amount of supersymmetry present in
five dimensions is then determined by the holonomy of
the internal space $K_{5}$.
One should also note that the constraint on $\eta$ is consistent with
the chirality on $\epsilon$. Defining the $10$-dimensional chirality
operator as
\bea
\Gamma^{11} = \Gamma^{0}\Gamma^{1}\cdots \Gamma^{9},
\eea
with $(\Gamma^{11})^{2} = 1$, we find that
\bea
\Gamma^{11} = iJ\gamma^{6}\otimes{\mathbf{1}}_{4}.
\eea
Hence,
\bea
\Gamma^{11}\epsilon = \epsilon \Rightarrow \;\;
iJ\gamma^{6}\eta = \eta.
\eea

\section{The Five-Dimensional anti-de Sitter Black Hole}
Recently, a higher-dimensional generalization of the BTZ
black hole was obtained \cite{Banados}. The construction is analogous
to the three-dimensional case, whereby a certain quotient of
anti-de Sitter space is constructed with the properties of a black
hole.
In particular, the five dimensional case was explicitly constructed.
The important point for our purposes is that the topology
of the black hole is ${\mathbf{R}}^{d-1}\times S^{1}$, with the horizon
being given by the $S^{1}$ factor. This is to be contrasted with
the topology ${\mathbf{R}}^{2}\times S^{d-2}$
of a Schwarzschild black hole.
The $\mathrm{adS}_{5}$ black hole is parametrized by two parameters,
its mass and angular
momentum, and as shown in \cite{Banados}, these can
be defined by relating the construction to
a Chern-Simons supergravity theory for the supergroup
$SU(2,2\mid N)$ \cite{Cham}.

The line element of the $\mathrm{adS}_{5}$ black hole
can be written in the form \cite{Banados}
\bea
ds^{2} &=& -\left(\frac{r^{2}-r_{+}^{2}}{\ell^{2}}\right)
\frac{\ell^{2}}{r_{+}^{2}}\cos^{2}\theta\; dt^{2}
+\left(\frac{r^{2} - r_{+}^{2}}{\ell^{2}}\right)^{-1}dr^2
+ r^{2}\frac{\ell^{2}}{r_{+}^{2}}\;d\phi^{2} \non\\
&+&(r^{2} - r_{+}^{2})\frac{\ell^{2}}{r_{+}^{2}}(d\theta^{2}
+ \sin^{2}\theta\;d\chi^{2}),
\label{ads5}
\eea
in the range $-\infty < t < \infty, r_{+} < r < \infty, 0< \theta < \pi,
0\leq \chi < 2\pi$. The location of the horizon is specified by
$r = r_{+}$.  We note that at the horizon the angular $(\theta,\chi)$
part of the line element also vanishes.
In order to introduce angular momentum, one makes
the replacements
\bea
t &\rightarrow&  \frac{r_{+}}{\ell}t - r_{-}\phi,\non\\
\phi &\rightarrow& \frac{r_{-}}{\ell^{2}}t - \frac{r_{+}}{\ell}\phi,
\eea
and identifies points along the new angular coordinate $\phi \sim
\phi + 2\pi n$, with $r_{-} < r_{+}$.
The line element then becomes
\bea
ds^{2} &=& dt^{2}\left[\frac{r^{2}r_{-}^{2}}{r_{+}^{2}\ell^{2}}
- \left(\frac{r^{2}-r_{+}^{2}}{\ell^{2}}\right)\cos^{2}\theta\right]
+\left(\frac{r^{2}-r_{+}^{2}}{\ell^{2}}\right)^{-1}dr^{2}
+d\phi^{2}\left[r^{2}-\left(\frac{r^{2}-r_{+}^{2}}{\ell^{2}}\right)
\frac{r_{-}^{2}\ell^{2}}{r_{+}^{2}}\cos^{2}\theta\right]\non\\
&+&dt\;d\phi\left[-\frac{2r^{2}r_{-}}{r_{+}\ell}
+\left(\frac{r^{2} - r_{+}^{2}}{\ell^{2}}\right)
\frac{2r_{-}\ell}{r_{+}}
\cos^{2}\theta\right]
+(r^{2}-r_{+}^{2})\frac{\ell^{2}}{r_{+}^{2}}(d\theta^{2}
+\sin^{2}\theta\; d\chi^{2}).
\eea

The Bekenstein-Hawking entropy for the $\mathrm{adS}_{5}$
black hole was computed in \cite{Banados}, and found to be
\bea
S_{BH} = 4\pi kr_{-},
\eea
where $k$ is the coupling parameter which multiplies
the Chern-Simons supergravity action.
Since this action is quadratic in curvature, it follows
that $k$ has dimensions $-1$ in length.
The mass and angular momentum
are given by
\bea
M = k\frac{2r_{+}r_{-}}{\ell^{2}},\;\;
J = k\frac{r_{+}^{2} + r_{-}^{2}}{\ell},
\eea
with the constraint that the horizon exists only for $J \geq M\ell$.
The interesting point, as observed in \cite{Banados}, is that
the entropy is not proportional to the length of $S^{1}$, namely
$2\pi r_{+}$.
However, we note that at extremality $ J = M\ell$, we
have $r_{+} = r_{-}$, and hence the Bekenstein-Hawking
entropy is given by
\bea
S_{BH} =  (2\pi r_{+}) 2k.
\eea
Due to the fact that the horizon for the five-dimensional
case is an $S^{1}$ factor, one notes that the
Bekenstein-Hawking entropy at extremality depends on the
square root of the mass, as in the case of the BTZ black hole.
For very massive BTZ black holes,
the entropy has been understood
at the microscopic level in \cite{Strominger2,BSS}. This understanding
is based on the special asymptotic properties of
$\mathrm{adS}_{3}$ \cite{Brown,CH}. Let us recall that in
the extreme case, the mass and entropy of the
BTZ black hole are given by \cite{BTZ}
\bea
M_{BTZ} = \frac{r_{+}^{2}}{\ell^{2}}\frac{1}{4G_{3}},\;\;
S_{BTZ} = \frac{2\pi r_{+}}{4G_{3}},
\eea
where $G_{3}$ is the three-dimensional Newton constant with
dimensions $+1$ in length.

In order to relate the mass and entropy of the $\mathrm{adS}_{5}$
black hole to the BTZ case, we need a relationship between
the three- and five-dimensional Newton constants.
Within string theory, it is natural to identify
$k = \alpha^{\prime}/8 G_{5}$, where
$G_{5}$ is the five-dimensional Newton constant with
dimensions $+3$ in length.
The  entropy is then given by
\bea
S_{BH} = (2\pi r_{+}) \frac{\alpha^{\prime}}{4G_{5}}.
\eea
However, in order of magnitude, we have
$G_{3} \sim G_{5}/\alpha^{\prime}$.
We thus see that the mass of the
five-dimensional black hole is of the order of the mass
of the BTZ black hole. Upon this identification, one then finds
that the entropy of the $\mathrm{adS}_{5}$ black hole is of the
order of the BTZ entropy (which has a statistical interpretation
\cite{Strominger2,BSS,Carlip2}).
In order words, this identification yields the correct mass dependence of
the five-dimensional entropy. In this sense, the analysis is
similar in spirit
to the correspondence principle presented in \cite{Susskind,HP}.
However, further work is required in order to fix the numerical
coefficient.

\section{Conclusions}
The construction of conserved charges for the five-dimensional
anti-de Sitter black hole relied on the formulation of
a Chern-Simons supergravity theory for the
supergroup $SU(2,2\mid N)$ \cite{Cham}.
In this regard, it is worth remarking that a computation of
the entropy for the BTZ black hole from the point of
view of three-dimensional Chern-Simons theory with boundary
was provided in \cite{Carlip2}.
The computation is based essentially on the observation that the
horizon dynamics on the boundary is controlled by a WZW model.
It would be interesting to see if a similar computation
in the $SU(2,2\mid N)$ Chern-Simons theory sheds further light
on the entropy.
Indeed, there has been recent progress in understanding the
connection between singleton field theories in
anti-de Sitter space and the D-brane picture of
black hole entropy \cite{Sfetsos}.
Further considerations are contained in \cite{Maldacena2}-\cite{Boon}.
One should also mention that the self-dual $3$-brane of
type IIB string theory has been discussed in the context of
vacuum interpolation between $10$-dimensional Minkowski space
and $\mathrm{adS}_{5}\times S^{5}$ \cite{Duff4,Gibbons}.

Another interesting observation is the fact that the
Chern-Simons supergravity theory for the anti-de Sitter
group can be constructed
up to a maximal dimension of seven \cite{Cham}.
One might therefore expect to be able to construct conserved
charges for the seven-dimensional anti-de Sitter black hole,
as in the five-dimensional case. Furthermore, the $\mathrm{adS}_{7}$
black hole could then be obtained as a compactification
$\mathrm{adS}_{7}\times K_{4}$ of
$11$-dimensional supergravity, with the $4$-form field strength
being proportional to the volume form of the compact internal
space $K_{4}$ \cite{FR,Duff3}. In this way, the $\mathrm{adS}_{7}$
black hole could be given an interpretation as a solution of
the low-energy limit of M Theory, with the $2$-brane
field excited.

\noindent {\large\bf Acknowledgements}\\
I would like to thank K. Sfetsos for valuable correspondence
on the entropy discussion.
This work was supported by Forbairt grant SC/96/603.


\begin{thebibliography}{99}
\bibitem{SV} A. Strominger and C. Vafa, {\em Microscopic Origin of the
Bekenstein-Hawking Entropy}, Phys. Lett. B379 (1996) 99,
hep-th/9601029.
\bibitem{Callan} C. Callan and J. Maldacena,
{\em D-brane Approach to Black
Hole Quantum Mechanics}, Nucl. Phys. B472 (1996) 591, hep-th/9602043.
\bibitem{Horowitz} See, for example, 
G. Horowitz, {\em Quantum States of Black Holes}, gr-qc/9704072;
J. Maldacena, {\em Black Holes in String Theory},
hep-th/9607235.
\bibitem{Polchinski} J. Polchinski, {\em Dirichlet-branes
and Ramond-Ramond Charges}, Phys. Rev. Lett. 75
(1995) 4724, hep-th/9510017.
\bibitem{BTZ} M. Ba\~nados, C.\ Teitelboim, and J.\ Zanelli, {\em Black
Hole in Three-Dimensional Spacetime},
Phys. Rev. Lett. 69 (1992) 1849, hep-th/9204099. 
\bibitem{BHTZ} M. Ba\~nados, M.\ Henneaux, C.\ Teitelboim, and J.\
Zanelli, {\em  Geometry of the $2+1$ Black Hole},
Phys. Rev. D48 (1993) 1506,
gr-qc/9302012.
\bibitem{Carlip} S. Carlip, {\em The $(2+1)$-Dimensional Black Hole},
Class. Quant. Grav. 12 (1995) 2853, gr-qc/9506079. 
\bibitem{Strominger2} A. Strominger, {\em Black Hole Entropy
from Near-Horizon Microstates}, hep-th/9712251.
\bibitem{Brown} J.D. Brown and M. Henneaux, {\em Central Charges
in the Canonical Realization of Asymptotic Symmetries: An Example
from Three Dimensional Gravity}, Commun. Math. Phys.
104 (1986) 207.
\bibitem{CH} O. Coussaert and M. Henneaux, 
{\em Supersymmetry of the $(2+1)$-Dimensional Black Holes},
Phys. Rev. Lett. 72 (1994) 183, hep-th/9310194.
\bibitem{BSS} D. Birmingham, I. Sachs, and S. Sen, {\em Entropy
of Three-Dimensional Black Holes in String Theory},
hep-th/9801019.
\bibitem{Banados} M. Ba\~{n}ados, {\em Constant Curvature
Black Holes}, gr-qc/9703040.
\bibitem{Duff2} M.J. Duff, R.R. Khuri, and J.X. Lu, {\em String
Solitons}, Phys. Rep. 259 (1995) 213, hep-th/9412184.
\bibitem{Schwarz} J.H. Schwarz, {\em Covariant Field Equations
of Chiral $N=2$ $D=10$ Supergravity}, Nucl. Phys. B226 (1983) 269.
\bibitem{Wetterich} C. Wetterich, {\em Dimensional Reduction
of Weyl, Majorana
and Majorana-Weyl Spinors}, Nucl. Phys. B222 (1983) 20.
\bibitem{Schwarz2} J.H. Schwarz, in {\em  Superstrings and Supergravity},
Proceedings of the 28th Scottish Universities Summer School in Physics,
ed. A.T. Davies and D.G. Sutherland, 1986.
\bibitem{Cham} A.H. Chamseddine, {\em Topological Gravity
and Supergravity in Various Dimensions}, Nucl. Phys. B346 (1990) 213.
\bibitem{Carlip2} S. Carlip, {\em Statistical Mechanics of the $(2+1)$-
Dimensional Black Hole}, Phys. Rev. D51 (1995) 632, gr-qc/9409052;
{\em The Statistical Mechanics of the Three-Dimensional Euclidean
Black Hole}, Phys. Rev. D55 (1997) 878, gr-qc/9606043.
\bibitem{Susskind} L. Susskind, {Some Speculations about
Black Hole Entropy in String Theory}, hep-th/9309145.
\bibitem{HP} G.T. Horowitz and J. Polchinski, {\em A Correspondence
Principle for Black Holes and Strings}, Phys. Rev. D55 (1997) 6189,
hep-th/9612146.
\bibitem{Sfetsos} K. Sfetsos and K. Skenderis, {\em Microscopic
Derivation of the Bekenstein-Hawking Entropy Formula for
Non-Extremal Black Holes}, hep-th/9711138.
\bibitem{Maldacena2} J. Maldacena, {\em The Large $N$ Limit of
Superconformal Field Theories and Supergravity},
hep-th/9711200.
\bibitem{Ferrara} S. Ferrara and C. Fronsdal, {\em Conformal
Maxwell Theory as a Singleton Field Theory
on $ADS_{5}$, IIB Three Branes and Duality}, hep-th/9712239.
\bibitem{Boon} H.J. Boonstra, B. Peeters, and K. Skenderis,
{\em Branes and anti-de Sitter spacetimes}, hep-th/9801076.
\bibitem{Duff4} M.J. Duff and J.X. Lu, {\em The Self-Dual
Type IIB Superthreebrane}, Phys. Lett. B273 (1991) 409.
\bibitem{Gibbons} G.W. Gibbons and P.K. Townsend,
{\em Vacuum Interpolation in Supergravity via Super $p$-branes},
Phys. Rev. Lett. 71 (1993) 3754, hep-th/9307049.
\bibitem{FR} P.G.O. Freund and M.A. Rubin, {\em Dynamics
of Dimensional Reduction}, Phys. Lett. 97B (1980) 233.
\bibitem{Duff3} M.J. Duff, B.E.W. Nilsson, and C.N. Pope,
{\em Kaluza-Klein Supergravity}, Phys. Rep. 130 (1986) 1.
\end{thebibliography}
\end{document}